\documentclass[12pt]{llncs}
\usepackage{graphicx,  amsmath, amssymb, algorithm, algpseudocode,caption} 

\title{On List Colouring and List Homomorphism of Permutation and Interval Graphs}

\author{Jessica Enright  \inst{1} \and Lorna Stewart \inst{1} \and G\'{a}bor Tardos \inst{2}}

\institute{University of Alberta\\ \and Simon Fraser University and R\'enyi Institute }

\date{}

\begin{document}
%
%
\newenvironment{my_enumerate}{
\begin{enumerate}
  \setlength{\itemsep}{1pt}
  \setlength{\parskip}{0pt}
  \setlength{\parsep}{0pt}}
  {\end{enumerate}
}
\newenvironment{my_itemize}{
\begin{itemize}
  \setlength{\itemsep}{1pt}
  \setlength{\parskip}{0pt}
  \setlength{\parsep}{0pt}}
  {\end{itemize}
}
\newenvironment{my_proof}{   
        \begin{proof} 
     }
     { $\Box$ \end{proof}
}

\maketitle

\begin{abstract}
List colouring is an NP-complete decision problem even if the total number of
colours is three.  It is hard even on planar bipartite graphs.  We give a
polynomial-time algorithm for solving list colouring of permutation graphs
with a bounded total number of colours. More generally we give a
polynomial-time algorithm that solves the list-homomorphism problem to any
fixed target graph for a large class of input graphs including all permutation
and interval graphs.
\end{abstract}

\section{Introduction}
A {\em proper colouring} of a graph assigns colours to the vertices such that adjacent vertices receive distinct colours. (In this paper we deal only with vertex colourings.) The {$k$-colouring problem} asks if a given graph has a proper colouring with at most $k$ colours. For $k\ge3$ this is NP-complete. 

In the {\em list colouring problem} each vertex of the input graph comes with a list of allowed colours and we ask if a proper colouring exists where each vertex receives a colour from its list. As a generalization of ordinary colouring, it is NP-complete \cite{toft}.  List colouring remains hard even on interval graphs \cite{biro1992}, as well as split graphs, cographs, and bipartite graphs \cite{jansenScheffler}.  It is solvable in polynomial time on trees \cite{jansenScheffler}.

Kratochv\'il and Tuza \cite{kratochvilTuza} showed that list colouring is NP-complete even if the size of each list assigned to a vertex is at most three, each colour appears in at most three lists, each vertex in the graph has degree at most three, and the graph is planar.  However, they gave polynomial-time algorithms to solve list colouring on a graph if the maximum list size is at most two, or each colour appears in at most two lists, or each vertex has degree at most two.

Let {\em $k$-list colouring} stand for the list colouring problem where the total number of colours is bounded by the constant $k$. This is a generalization of $k$-colouring, thus for $k\ge3$ it is NP-complete.  It remains NP-complete on planar bipartite graphs \cite{kratochvilFixedColourBound}, but is solvable in polynomial time on graphs of fixed treewidth \cite{bigPaper}.    

Note that $2$-list colouring is solvable in polynomial time. Indeed,
2-colouring is solvable in polynomial time and has at most two
(complementary) solutions on each connected component. Thus, for the
$2$-list colouring problem it is enough to check that one of these is
compatible with the lists on each component.

A {\em graph homomorphism} from a graph $G$ to another graph $H$ is a
function $f:V(G)\to V(H)$ satisfying that $f(x)$ and $f(y)$ are adjacent in
$H$ whenever $x$ and $y$ are adjacent in $G$. Note that here we allow the
graphs to have loops.

Let $H$ be fixed graph. The $H$-colouring problem takes a graph $G$ as
input and asks if there is $G$ to $H$ homomorphism. In the list $H$-colouring
problem each vertex of the input graph comes with a list of vertices of $H$
and we ask if a $G$ to $H$ homomorphism exists that maps each vertex to a
member of its list. Clearly, $k$-colouring is a
graph-homomorphism to the complete graph $K_k$, so list $H$-colouring is a
generalization of $k$-list colouring.

Permutation graphs are comparability cocomparability graphs (see definitions
in the next section). List colouring is NP-complete
on permutation graphs since cographs are
permutation graphs \cite{jansenScheffler}. The $k$-list colouring problem is
NP-complete for comparability graphs for $k\ge3$, since bipartite graphs are
comparability graphs. The complexity of $k$-list colouring of cocomparability
graphs remains open.

In this paper we give a polynomial-time algorithm for the $k$-list colouring
of permutation graphs for any fixed $k$. More generally we give a
polynomial-time algorithm that solves the list-homomorphism problem to any
fixed target graph for permuatation graphs. The same algorithm also works for
interval graphs and more.

Our algorithm is based on what we call a {\em multi-chain
ordering} (see definition in the next section), a notion related to 
chain graphs \cite{Yann} and to a characterization of 
bipartite permutation graphs given in \cite{BL}.
The algorithm applies to every graph with all connected induced subgraphs
having a multi-chain ordering, among them all permutation graphs and
all interval graphs. We also remark that since adding loops to a graph does not have
any effect on the multi-chain ordering, our algorithm also applies to interval
and permutation graphs with loops added to some vertices.
The running time for $k$-list colouring, or more
generally, for list $H$-colouring for a graph $H$ on $k$ vertices is
$O(n^{k^2-3k+4})$, where $n$ stands for the number of vertices of the input
graph.

 Ho{\`{a}}ng \textit{et al.} \cite{hoangP5} give an algorithm for $k$-list colouring 
 $P_5$-free graphs in polynomial time. 
 Their algorithm, like ours, is based on how a colouring of one side of a bipartition of a chain graph
 can restrict the other side to a polynomial number of possible colourings.  
 


We mention here that a polynomial-time $k$-list colouring algorithm for
interval graphs cannot be considered
new. Indeed, another polynomial-time algorithm already exists for list
$H$-colouring graphs
with bounded treewidth. The treewidth of an interval graph is
one less than the size of its largest clique. Thus, unless it is
bounded one does not have a proper colouring with a bounded number of
colours. The same cannot be said about permutation graphs though. Even
bipartite permutation graphs have unbounded treewidth.

Multi-chain orderings are based on distance from a
starting vertex. They give insight into the structure of permutation
or interval graphs, and may lead to algorithms for other problems on
these or similar graphs.

\section{Definitions and Preliminaries}
We consider finite graphs only with no multiple edges. We allow for loop
edges connecting a vertex to itself and call a graph {\em simple} if it has
no such edge. (Loop edges in the input graph only make
sense for the list
$H$-colouring problem if $H$ has at least one loop, so in particular, not for
$k$-list colouring.) We represent graphs as a pair $G = (V, E)$, where
$V=V(G)$ is the vertex set and $E=E(G)$ is the edge set. We denote the edge
connecting $x$ to $y$ by $xy$, so $xy=yx$. In a {\em
directed graph} we have ordered pairs of vertices as edges and
denote such an edge as $\overrightarrow{uv}$ saying it leaves the
vertex $u$ and is oriented toward the vertex $v$. A {\em sink} is a
vertex that no edge leaves. A directed graph is {\em transitive} if
the presence of the edges $\overrightarrow{uv}$ and $\overrightarrow{vw}$
implies the presence of $\overrightarrow{uw}$. An {\em orientation} of
the simple graph $G=(V,E)$ is a directed graph $G=(V,\overrightarrow E)$,
where $\overrightarrow E$ is obtained by replacing each edge
$\{u,v\}\in E$ by one of its orientations: $\overrightarrow{uv}$ or
$\overrightarrow{vu}$ but not both. A {\em comparability graph} is a
simple graph that admits a transitive orientation. Equivalently, a graph is a
comparability graph if there is a partial order on the vertices with
exactly the adjacent (distinct) vertices being comparable. The {\em
complement} of the simple graph $G=(V,E)$ is $\overline G=(V,\overline E)$,
where $\overline E$ contains all possible non-loop edges on $V$ not in $E$. We
sometimes call the edges of $\overline G$ the {\em nonedges} of $G$. A {\em
cocomparability graph} is a graph whose complement is a
comparability graph.
Graphs that are simultaneously comparability and cocomparability graphs are called {\em permutation graphs}. Permutation graphs are exactly the graphs $G=(V,E)$ that are obtained from a permutation $\pi:\{1,\ldots,n\}\to\{1,\ldots,n\}$ by setting $V=\{x_1,\ldots,x_n\}$ and $E=\{x_ix_j\mid i<j,\pi(i)<\pi(j)\}$. A
simple graph is an {\em interval
graph} if one can identify its vertices with real intervals
such that two vertices are adjacent if and only if the corresponding
intervals intersect.
Such intervals can always be chosen to have distinct endpoints.
{\em Weakly chordal graphs} are simple graphs with no induced $C_n$ or $\overline{C_n}$, for $n>4$.

Let $G = (V, E)$ be a graph.  A \emph{list mapping} of $G$ is a
mapping that assigns a set (list) of colours to each vertex in $G$.
A colouring of $G$ \emph{obeys} a list mapping if it assigns every
vertex a colour from its list. More generally, if the graph $H$ is fixed a
{\em list mapping} of $G$ assigns a subset of $V(H)$ (a list) to every vertex
of $G$. A homomorphism from $G$ to $H$ {\em obeys} the list mapping if each
vertex is mapped to member of its list.

A {\em chain graph} is a bipartite graph that contains no induced $2K_2$.
This name was introduced by Yannakakis \cite{Yann}. The following
characterization is easily seen to be equivalent to the definition. A
bipartite graph with sides (partite sets) $A$ and $B$ is a chain graph if and
only if for any two vertices in $A$ the neighborhood of one of them contains
the neighborhood of the other. As a consequence we see that if we order the
vertices of $B$ according to decreasing degree (breaking ties arbitrarily),
then the neighborhood of any vertex in $A$ consists of 
a consecutive (in the ordering) set of vertices in $B$, including the first vertex of $B$.

Let $G = (V, E)$ be a connected graph. The {\em distance layers of
$G$} from a vertex $v_0$ are $\{v_0\}=L_0,L_1,\ldots,L_z$, where $L_i$
consists of the vertices at distance $i$ from $v_0$ and $z$ is the
largest integer for which this set is not empty. These layers form
a \emph{multi-chain ordering} of $G$ if for every two consecutive layers $L_i$
and $L_{i+1}$ the edges connecting these two layers form a chain graph.

Our algorithm processes each connected component of the input graph
separately. It is based on multi-chain orderings of the components and uses the
following simple properties of such orderings: (a) $H$-colouring of one
layer in a multi-chain ordering has limited effect on the colouring
of the next layer and no direct effect on subsequent layers and (b)
each layer has a vertex that is adjacent to all vertices in the next
layer, thus if this vertex is mapped to $c$ then all non-neighbours of $c$ will be
missing from the $H$-colouring of the next layer, practically reducing the
size of $H$. Note that (b) does not apply if $H$ has a vertex $c$ that is
adjacent to every vertex of $H$ including itself. Fortunately this easy
special case can be handled by alternate methods.

\begin{lemma}\label{lem:TransReg}
Let $\overrightarrow{G}=(V,\overrightarrow{E})$ be a transitive
orientation of a connected comparability graph $G=(V,E)$. Let $v_0\in V$ be a
sink 
in $\overrightarrow G$ and let $L_0,\ldots,L_z$ be the distance layers
of $G$ from $v_0$.
Then for $0\le i<z$ all edges of $\overrightarrow{E}$ between the
vertices of two consecutive layers $L_i$ and $L_{i+1}$ are oriented
toward $L_i$ if $i$ is even and all these edges are oriented toward
$L_{i+1}$ if $i$ is odd.
\end{lemma}
\begin{my_proof}
We proceed by induction on $i$. For $i=0$ the statement of the lemma
holds since $v_0$ is a sink. Each $u\in L_i$ for $i>0$ has a neighbour
$u'\in L_{i-1}$, and an edge between $u$ and $L_{i+1}$ oriented
``the wrong way'' would imply the presence of an edge between $u'$ and
$L_{i+1}$ by transitivity, a contradiction.
\end{my_proof}

\begin{lemma}\label{lem:TransComp}
Let $\overrightarrow{\overline{G}}=(V,\overrightarrow{\overline{E}})$
be a transitive orientation of the complement of a connected comparability graph
$G = (V, E)$. Let  $v_0\in V$ be a sink in $\overrightarrow{\overline
G}$ and let $L_0,\ldots,L_z$ be the distance layers of
$G$ from the vertex $v_0$. Then for
every pair of layers $L_i, L_j$ where $0 \leq i < j \leq z$ all edges
of $\overrightarrow{\overline G}$ between $L_i$ and $L_j$ are directed
toward $L_i$.
\end{lemma}
\begin{my_proof}
We proceed by induction on $i$.  For $i=0$ the statement follows from
$v_0$ being a sink. Let us consider $i>0$ and assume for contradiction
that $\overrightarrow{uv}$ is an edge of $\overrightarrow{\overline G}$
with $u\in L_i$ and $v\in L_j$, $j>i$. Now $v$ is not adjacent in $G$
to any vertex $u'\in L_{i-1}$, so by the induction hypothesis we have
$\overrightarrow{vu'}\in\overrightarrow{\overline E}$ and by transitivity
$\overrightarrow{uu'}\in\overrightarrow{\overline E}$. But this
contradicts the fact that $u$ has a neighbour $u'\in L_{i-1}$ in
$G$, so no orientation of an edge between $u$ and this neighbour
should be present in $\overrightarrow{\overline G}$.
\end{my_proof}

\begin{theorem}\label{perm}
Every connected permutation graph has a multi-chain ordering.  
\end{theorem}
\begin{my_proof}
Let $G = (V, E)$ be a permutation graph and let $\overrightarrow{G}$
be a transitive orientation of $G$ and $\overrightarrow{\overline{G}}$
a transitive orientation of the complement of $G$.

Let $v_0$ be a vertex that is a sink in both
of the graphs $\overrightarrow G$ and $\overrightarrow{\overline G}$,
the existence of which is shown in \cite{pnueli}.
We claim that the distance layers $L_0,\ldots,L_z$ of $G$ from
$v_0$ form a multi-chain ordering.  
To see this assume for a contradiction that $u, v\in L_i$
and $u',v'\in L_{i-1}$ are vertices of two neighbouring layers and
$u$ is adjacent with $u'$ but not with $v'$
in $G$ and $v$ is adjacent with $v'$ but not with $u'$.  We
distinguish two cases according to whether $u$ and $v$ are adjacent in
$G$.

Assume first that $u$ and $v$ are adjacent and assume without loss of
generality that this edge is oriented
toward $v$ in $\overrightarrow G$.
By Lemma \ref{lem:TransReg}, either both the edge between $u$ and $u'$
and between $v$ and $v'$ are oriented toward $L_i$, or both are
oriented toward $L_{i-1}$.  In the former case we should have
$\overrightarrow{u'v}\in \overrightarrow E$ by transitivity,
contradicting our assumption that $v$ is not adjacent with $u'$ in
$G$. In the latter case we similarly have $\overrightarrow
{uv'}\in\overrightarrow E$, again a contradiction.

Now assume that $u$ and $v$ are not adjacent in $G$ and assume
without loss of generality that this nonedge
is oriented toward $v$ in $\overrightarrow{\overline G}$. By
Lemma \ref{lem:TransComp}, the nonedge between $v$ and $u'$ is
oriented toward $u'$ in $\overrightarrow{\overline G}$. By
transitivity we have $\overrightarrow{uu'}\in\overrightarrow{\overline
E}$ contradicting our assumption that $u$ and $u'$ are adjacent in $G$.
\end{my_proof}

Not every graph with a multi-chain ordering is a
permutation graph. Further examples are given by interval graphs as shown by
the next theorem. 
In addition, $C_n$ and $\overline{C_n}$ where $n>4$, and the graph $T$ defined as $K_{1,3}$ with each edge subdivided once, do not have multi-chain orderings.
Therefore, there are cocomparability graphs and even trees that do not have multi-chain orderings. Moreover, any graph such that all induced subgraphs have multi-chain orderings must be a weakly chordal graph, but not all weakly chordal graphs have multi-chain orderings. We also note that the complement of the graph $T$ is a cocomparability graph that is neither a permutation graph nor an interval graph, in which every connected subgraph has a multi-chain ordering. For further information about these graph classes, the reader is referred to \cite{BLS}.

\begin{theorem}\label{inter}
All connected interval graphs admit multi-chain orderings.
\end{theorem}

\begin{my_proof}
Consider an interval representation in which all interval endpoints are distinct.
We can choose $v_0$ to be the vertex with the leftmost left endpoint.
One can find reals $x_0<x_1\ldots$ such that the layer $L_i$ of $G$ at
distance $i$ from $v_0$ consists of the vertices with left endpoint
in $(x_{i-1},x_i]$. To see that these layers form  a multi-chain
ordering of $G$ take two vertices (intervals) in $L_i$ and let $u$ be the one with
its left end point more to the left, and $v$ the other. Clearly, all
intervals in $L_{i-1}$ intersecting $v$ must also intersect $u$.
\end{my_proof}

Our list $H$-colouring algorithm works for every graph whose
connected induced subgraphs all have multi-chain orderings, and it runs in
polynomial time as long as $H$ is fixed. The last two theorems show that
this class includes all permutation and interval graphs (and the graphs
obtained from them by adding some loops). Restricting attention to complete
graphs $H=K_k$ we get polynomial-time $k$-list colouring algorithms. 

Given a connected graph $G$ and vertex $v$ of $G$,
we can check whether the distance layers from starting vertex $v$
form a multi-chain ordering in $O(m)$ time where $m$ is the number of edges
of $G$.
The algorithm for doing so uses breadth-first search to
generate the distance layers from $v$ and to compute the degree
of each vertex in the next layer.
It then uses bucket sort to order the vertices of each layer
by decreasing size of their neighbourhood in the previous layer.  
Finally it checks that for each vertex it holds that its neighbours in the
next layer appear in the beginning of that layer before the
non-neighbours. Each step can be accomplished in $O(m)$ time. 

 As a naive algorithm to check if a connected graph has a multi-chain ordering,
 and generate it if it does, we can start a breadth-first search from
 each vertex, and check to see if that search has given us a
 multi-chain ordering in $O(nm)$ time overall.  In some classes,
 for example permutation graphs, this can be done more quickly.  In
 the case of permutation graphs, we can use the output of the linear-time
 recognition algorithm provided by McConnell and Spinrad
 \cite{mcconnell} to  identify a vertex that is a sink in some
 transitive orientation of both the graph and its complement. We can
 then generate the distance layers from this vertex in $O(m)$ time which is
 a multi-chain ordering as the proof of Theorem~\ref{perm} shows. Similarly,
several linear time algorithms exist to find a ``leftmost'' vertex in a interval
 graph, the earliest one being \cite{BoL}. The distance layers can be constructed from there
 in linear time. As the proof of
 Theorem~\ref{inter} shows this is a multi-chain ordering.

\section{The algorithm}
In this section we present our algorithm to list $H$-colour any graph with
the property that all connected induced subgraphs have multi-chain
orderings. The algorithm runs in polynomial time if $H$ is fixed.
Since the algorithm handles connected components separately,
we consider only connected graphs in the following description.

Let $G=(V,E)$ be a connected graph and let $L_0,\ldots,L_z$ form a multi-chain
ordering of $G$. For $x\in L_i$ we introduce $d_-(x)$ for the number
of neighbours of $x$ in $L_{i-1}$ (or $0$ if $i=0$) and $d_+(x)$ for
the number of neighbours of $x$ in $L_{i+1}$ (or $0$ if $i=z$). We fix an
ordering of the vertices within each layer according to decreasing $d_-$
values breaking ties arbitrarily. 
As observed in the definition of chain graphs, this ordering ensures that the
neighbours of a vertex $x\in L_i$ among the vertices of the next layer
$L_{i+1}$ must be the first $d_+(x)$ vertices in that layer.

Let us fix the target graph $H$ with vertex set $C=V(H)$.
Let $\mathcal P$ be a list mapping of $G$, so $\mathcal P(x)\subseteq C$ for
every vertex $x\in V$.

A \emph{configuration} is a pair $(i,B)$, where $1\le i\le z$ and
$B:C\to\{0,1,\ldots,|L_i|\}$ satisfying that $B$ takes both $0$ and
$|L_i|$ as values. We introduce two more special configurations:
$S_0=(0,B_0)$ and $S_{z+1}=(z+1,B_0)$, where $B_0:C\to\{0\}$ is the constant
zero function.

These configurations form the vertices of the {\em configuration
graph}. This is a directed graph that contains the edge from
$(i,B)$ to $(i',B')$ if $i'=i+1$ and there
is a homomorphism $\chi$ from the subgraph $G_i$ of $G$ induced by the
layer $L_i$ to $H$ {\em providing} for this edge, i.e.,
satisfying the following three conditions:

\begin{itemize}
\item {$\chi$ obeys $\cal P$, i.e., for $x\in L_i$ we have
  $\chi(x)\in\mathcal P(x)$.}
  
\item{$\chi$ does not assign $c\in C$ to the first $B(c)$
  vertices in $L_i$ (recall that $L_i$ is ordered).}
  
\item {For each $x\in L_i$ and $c\in C$ with $c$ not adjacent
  to $\chi(x)$ in $H$ we have $B'(c)\ge d_+(x)$.}
\end{itemize}

We call a vertex of the graph $H$ \emph{universal} if it is connected to every
vertex of $H$. In particular, a universal vertex must be connected to itself
too. The importance of the configuration graph is shown by the following
theorem.

\begin{theorem}\label{path}
Assume $H$ has no universal vertex. Then $G$ has a homomorphism to $H$ obeying
$\cal P$ if and only if there exists a directed path from $S_0$ to $S_{z+1}$
in the configuration graph.
\end{theorem}

\begin{my_proof}
Assume $\chi:V\to C$ is a homomorphism from $G$ to $H$ obeying $\cal P$. For
$1\le i
\le z$ define the function $B_i$ on $C$ by setting $B_i(c)$ to be the largest
integer $0\le B_i(c)\le|L_i|$ satisfying that $\chi$ does not map any of the
first $B_i(c)$ vertices of $L_i$ to $c$. Clearly, $B_i$ takes the value $0$ on
$\chi(x)$ for
the first vertex $x$ of $L_i$. We know that the vertices in the layer $L_i$
have a common neighbour $y$ in $L_{i-1}$. As $\chi(y)$ is not universal in $H$
there must exist $c\in C$ not adjacent to $\chi(y)$ and thus $\chi$ cannot
take the value $c$ on any neighbour of $y$ making $B_i(c)=|L_i|$. Thus
$S_i=(i,B_i)$ is a configuration. We
claim that $S_0S_1\ldots S_zS_{z+1}$ is a directed path in the
configuration graph. Indeed, for $0\le i\le z$ the restriction of
$\chi$ to $L_i$ provides for the edge
$\overrightarrow{S_iS_{i+1}}$. Conditions (1) and (2) are satisfied
trivially; to see (3) one has to use our observation that the
neighbours in $L_{i+1}$ of any vertex $x\in L_i$ are the first
$d_+(x)$ vertices of that layer.

Conversely, let us assume that there is a directed path from $S_0$ to
$S_{z+1}$ in the configuration graph. By the layered structure of the
configuration graph this path must be of the form
$S_0S_1\ldots S_zS_{z+1}$ with $S_i=(i,B_i)$ and appropriate
functions $B_i$. For $0\le i\le z$ let $\chi_i:L_i\to C$ be a homomorphism
providing for the $\overrightarrow{S_iS_{i+1}}$ edge and let
$\chi: V\to C$ be the union of these maps. We claim that $\chi$ is a $G$ to
$H$ homomorphism obeying $\cal P$.

The function $\chi$ obeys $\cal P$ since all its parts $\chi_i$ do so by
condition (1).

To see that $\chi$ is a homomorphism we have to show that the image of every edge
$xy\in E$ is an edge in $H$. Clearly, $x$ and $y$ have to come from the same
or neighbouring layers. If they are in the same layer $L_i$, then
$\chi(x)\chi(y)=\chi_i(x)\chi_i(y)\in E(H)$ because $\chi_i$ is a
homomorphism. Now assume that for some $0\le i<z$ we have
vertices $x\in L_i$ and $y\in L_{i+1}$ such that their images
$\chi(x)=\chi_i(x)$ and $\chi(y)=\chi_{i+1}(y)$ are not adjacent in $H$. By
condition (2) $\chi_{i+1}$ does not map the first $B_{i+1}(\chi(y))$ vertices
of $L_{i+1}$ to $\chi(y)$. Thus $y$ is not among the first $B_{i+1}(\chi(y))$
vertices of $L_{i+1}$. By condition (3) on $\chi_i$ we have
$B_{i+1}(\chi(y))\ge d_+(x)$, so $y$ is not among the first $d_+(x)$ vertices
of $L_{i+1}$, so it is not adjacent to $x$ as needed.
\end{my_proof}

Our next theorem tells us how to construct the configuration graph,
more precisely, how to decide whether an edge is present. Let us fix
two configurations $S=(i,B)$ and $S'=(i+1,B')$. Let $G_i$ be
the subgraph of $G$ induced on the layer $L_i$ and let us define a
list mapping $\mathcal P'$ on $G_i$ as follows. For $1\le j\le|L_i|$
let $x_j$ stand for the $j$'th vertex in the layer $L_i$ and let us
set $\mathcal P'(x_j)=\{c\in\mathcal P(x_j)\mid B(c)<j, \forall c'\in C(d_+(x_j)\le
B'(c')\hbox{ or }cc'\in E(H))\}$.

\begin{theorem}\label{const}
With $S$, $S'$, $G_i$ and $\mathcal P'$ as above there is an edge from
$S$ to $S'$ in the configuration graph if and only if $G_i$ has a homomorphism
to $H$ obeying $\mathcal P'$.
\end{theorem}

\begin{my_proof}
Any homomorphism providing for $\overrightarrow{SS'}$ obeys $\mathcal P'$ by
the conditions (1--3). Conversely any homomorphism from $G_i$ to $H$ that
obeys $\mathcal P'$ provides for this edge.
\end{my_proof}

We now present our algorithm for the list $H$-colouring problem for graphs with
all connected induced subgraphs having a multi-chain ordering. 

\begin{algorithm}
\caption{LH($G$, $\cal P$, $H$)}
\begin{algorithmic}

\smallskip

\State {\bf Input}: Graphs $G$, $H$, list mapping $\cal P$ where every connected
induced subgraph of $G$ must have a multi-chain ordering
\State {\bf Output}: TRUE if there is a homomorphism from $G$ to $H$ obeying
$\cal P$; FALSE otherwise
\smallskip
\State Let $H'$ be the subgraph of $H$ induced by vertices that appear in at
least one list of $\mathcal P$.

\If {$H'\ne H$}
  \Return LH($G$, $\cal P$, $H'$)
\EndIf

\If {$H$ has a universal vertex $c$}
  \State Let $G'$ be the subgraph of $G$ induced by the vertices $x$ with
  $c\notin\mathcal P(x)$
  \State \hspace{\algorithmicindent}  and let $\mathcal P'$ be the restriction of $\mathcal P$
  to this subgraph.
  \Return LH($G'$,$H$,$\mathcal P'$)
\EndIf

\If {$G$ has a single vertex}
  \If {$H$ has a loop or $G$ has no loop and $H$ has at least one vertex}
    \Return TRUE
  \Else
    \Return FALSE
  \EndIf
\EndIf

\For {each connected component $D=(V,E)$ of $G$}

  \If {$H$ has at most two vertices}
      \State Find all (the at most two) homomorphisms from $D$ to $H$.
       \If {at least one of the homomorphisms obeys $\mathcal P$}
            \State $c_D \leftarrow$ TRUE
        \Else
                 \State $c_D \leftarrow$ FALSE
        \EndIf

    \Else
        \State Find a multi-chain ordering $L_0,\ldots,L_z$ of $D$
        and order the vertices of 
        \State \hspace{\algorithmicindent} each layer by decreasing size of
        neighbourhood in the next layer.
        
        \State Initialize the directed configuration graph to have a vertex
        for each 
        \State \hspace{\algorithmicindent} configuration of this multi-chain ordering including $S_0$
        and $S_{z+1}$.

        \For {$i \leftarrow 0$ {\bf to} $z-1$}
        \State Let $D_i$ be the subgraph of $D$ induced by $L_i$.
             \For {each pair of configurations $S = (i,B)$ and $S' = (i+1,B')$}
                \State Construct a list mapping $\cal P'$ for $D_i$ as follows.
                \For {$j \leftarrow 1$ {\bf to} $|L_i|$}
                  \State Let $x_j$ stand for the $j$'th
                vertex in the layer $L_i$.
                  \State $\mathcal P'(x_j) \leftarrow 
                                \{c\in\mathcal P(x_j)\mid B(c)<j, $
                                \State \hspace{\algorithmicindent}  $\forall
                                c'\in V(H)(d_+(x_j)
                               \le B'(c')\hbox{ or
                                }cc'\in E(H))\}$
                \EndFor
                \If {LH($D_i$, $\mathcal{P'}$, $H$) = TRUE}
                      \State Add edge $\overrightarrow{SS'}$ to the
                      configuration graph
                \EndIf
             \EndFor
        \EndFor
                 \algstore{myalg}
  \end{algorithmic}
\end{algorithm}
\clearpage

\begin{algorithm}
  \ContinuedFloat
  \caption{LH($G$, $\cal P$, $H$) (continued)}
  \begin{algorithmic}
      \algrestore{myalg}
        \If {there is a directed path from $S_0$ to $S_{z+1}$ in the configuration graph}
         \State $c_D \leftarrow$ TRUE
        \Else 
                 \State $c_D \leftarrow$ FALSE
        \EndIf
\EndIf
\EndFor

\If {$c_D =$ TRUE for all components $D$ of $G$}
\Return TRUE
\Else
\Return FALSE
\EndIf
\end{algorithmic}
\end{algorithm}

We are given a fixed graph $H$, an input graph $G$ and the list mapping
$\mathcal P$. We start with very simple reductions.

If $H$ has a universal vertex $c$, then consider the subgraph $G'$ of $G$
induced by the vertices whose lists do not contain $c$. Clearly, $G$ has a
homomorphism to $H$ obeying $\mathcal P$ if and only if $G'$ has such a
homomorphism as the vertices outside $G'$ can ``freely'' be mapped to $c$.

Let $H'$ stand for the subgraph of $H$ induced by all the vertices that appear
in the lists in $\mathcal P$. Clearly, $G$ has a homomorphism to $H$ obeying
$\mathcal P$ if and only if $G$ has a homomorphism to $H'$ obeying $\mathcal
P$.

$G$ has homomorphism to $H$ obeying $\mathcal P$ if and only if all connected
components of $G$ have homomorphisms to $H$ obeying $\mathcal P$.

We use the these reductions (repeatedly, if necessary) until we arrive at a
problem in which $G$ is connected, $H$ has no universal
vertex and each vertex of $H$ appears on a list of $\mathcal P$.

Start by constructing the layers $L_0,\ldots,L_z$ of a multi-chain ordering of
$G$ with the corresponding ordering of the vertices within the layers
according to decreasing $d_-$ degrees. Construct the
configurations for this multi-chain ordering including $S_0$ and $S_{z+1}$.
Construct the edges of the configuration graph using a recursive
  call to check for the presence of each possible edge using the equivalent
  condition as given in Theorem~\ref{const}.
Return TRUE if there is directed path from $S_0$ to
  $S_{z+1}$ in the configuration graph and
  return FALSE otherwise.

Note that the recursive calls to determine the presence of an edge
from the configuration $(i,B)$ to $(i+1,B')$ is simpler than the original
problem instance. Indeed, it is a list $H$-colouring problem for $G_i$ and
$G_i$ has a single vertex for $i=0$, while for $i>0$ we have a vertex $c$ of
$H$ with $B(c)=|L_i|$ and this vertex does not show up in any of the lists ---
basically decreasing the number of vertices in the target graph $H$. To give
base to this recursion we solve the trivial instances directly:
If either $G$ or $H$ has a single vertex, deciding the list $H$-colouring
problem for $G$ becomes trivial. We can also handle the case where $H$ has
two vertices directly. If the two vertices are not adjacent in $H$, we must
map each connected component of $G$ to one or the other vertex. If the two
vertices of $H$ are connected and there is no loop in $H$ we face a $2$-list
colouring problem already discussed in the introduction. Finally if the two
vertices of $H$ are connected and there is also a loop in $H$, then $H$ has a
universal vertex and list $H$-colouring reduces to list $H'$-colouring with
$H'$ having a single vertex.

Using Theorems~\ref{path} and \ref{const} it is
straightforward to see that the above algorithm correctly answers the
question whether $G$ has a homomorphism to $H$ obeying $\cal P$.

It is a bit more involved to estimate the running time. Let $k$ and $n$ stand
for the number of vertices in $H$ and $G$. We claim the the running time of the
algorithm is $O(n^{k^2-3k+4})$ (the constant of proportionality depends on
$k$). We prove
this statement by induction on $k$. For $k\le 2$ the algorithm clearly
finishes in time $O(n^2)$. Let us assume $k>2$. If $H$ has a universal
vertex our reduction reduces list $H$-colouring to a single list $H'$-colouring
instance with $H'$ having fewer vertices. If $H$ has no universal vertex we
split $G$ into connected components, find the multi-chain ordering of each
component and build the configuration graphs corresponding to them. The
number of configurations for a fixed layer $L_i$ of a single component is
$O(|L_i|^{k-2})$ because the value of the function $B$ in a configuration
$(i,B)$ is arbitrary for $k-2$ vertices of $H$, but it has to be either $0$ or
$|L_i|$ for two. So the number of configurations for all connected components
together can be bounded by $O(n^{k-2})$ and the number of potential edges (the
number of recursive calls on the top level) is $O(n^{2k-4})$. In a
recursive call to test the presence of an edge in the configuration graph one
uses a list mapping that avoids at least one vertex of $H$ completely, so the
inductive hypothesis can be used for $k-1$. The only exception to this
rule is the test for an edge leaving the configuration $S_0$ of one of the
components,  but there the recursive call is for a
trivial graph on $|L_0|=1$ vertices. These trivial recursive calls
take constant time, the other recursive calls take
$O(n^{(k-1)^2-3(k-1)+4})$ time, so all recursive calls finish in
$O(n^{2k-4}n^{(k-1)^2-3(k-1)+4})=O(n^{k^2-3k+4})$ time. This huge time bound
clearly dominates the time of the non-recursive part of the algorithm.

\section{Conclusion}
We have given a polynomial-time algorithm to solve the list $H$-colouring
problem for fixed $H$ if every connected induced subgraph of the input graph
has a multi-chain ordering. Every connected permutation or interval graph has
a multi-chain ordering, so this algorithm works for permutation and interval
graphs.

\bibliography{listColouring}

\end{document}